\documentclass[journal=jacsat,manuscript=article]{achemso}

\usepackage[version=3]{mhchem}
\usepackage{makecell}
\usepackage{multirow}
\usepackage{graphicx}

\author{Khondokar Zahin}
\author{A.K.M. Hasibul Hoque}
\author{Md. Jawadul Karim}
\affiliation[Unknown University]
{Department of Electrical and Electronic Engineering, Bangladesh University of Engineering and Technology, ECE Building, West Palashi Campus, Dhaka 1205, Bangladesh}
\author{Ying Yin Tsui}
\affiliation[Unknown University]
{Department of Electrical and Computer Engineering, University of Alberta, Edmonton, AB T6G 2H5 Canada}
\author{Md Zahurul Islam}
\affiliation[Unknown University]
{Department of Electrical and Electronic Engineering, Bangladesh University of Engineering and Technology, ECE Building, West Palashi Campus, Dhaka 1205, Bangladesh}
\email{mdzahurulislam@eee.buet.ac.bd}

\title[An \textsf{achemso} demo]
  {Inverse Design Optimization of MIS Nanorod Array to Enhance Wavelength-Selective Light Extraction of OLED through Guided SPCE}

\abbreviations{IR,NMR,UV}

\begin{document}


\begin{abstract}
One of the primary causes of low light extraction efficiency (LEE) in conventional OLEDs is the trapping of light within the substrate layer. To overcome this limitation, various nanostructures have recently been investigated for integration into different OLED layers, aiming to convert the trapped loss modes into out-coupled light and thereby improve LEE. In this work, we propose a simple yet innovative nanorod array design composed of metal, insulator, and semiconductor-oxide layers, integrated on the bottom surface of the OLED structure. This configuration enhances light extraction through guided surface plasmon-coupled emission (SPCE). Using an inverse design-based optimization approach, the structural parameters of the nanorods were optimized to achieve wavelength-selective LEE enhancement. The periodic MIS nanorod array was optimized via the Particle Swarm Optimization (PSO) algorithm for different emission wavelengths, yielding improvements of 29\%, 30\%, and 51\% at 440 nm, 550 nm, and 540 nm, respectively, compared with conventional OLEDs. In addition, the optimized arrays demonstrate high spectral luminous efficiency (SLE), reduced UV transmittance, and uniform emission directionality. These enhancements have significant potential to advance OLED-based display technologies.
\end{abstract}

\section{Introduction}
Organic Light-Emitting Diodes (OLEDs) are thriving in the field of display technologies due to their capability of producing natural colors, low power consumption, fast response, and many other advantages. For high-efficiency light emission, an electron transport layer (ETL), a hole transport layer (HTL), and electron and hole blocking layers have been introduced to enhance the exciton recombination rate in the emission layer. Variation in thickness of the layers, such as ETL and HTL, has been studied for enhancing LEE of OLED\cite{Lin:22, ULLA2024114602}. Micro-patterned low-index grid to extract waveguide modes, anti-reflection layer to reduce total internal reflection (TIR), and deposition of external hierarchical texture on bottom-emitting OLED to improve external quantum efficiency (EQE) have been common in OLED structures\cite{chen2010high,sun2008enhanced, Kovacic:21}. Despite these modifications, progress in OLED performance has stagnated, which demands the emergence of nanostructures in these devices. There are several types of nanorstructures integrated into OLEDs, such as periodic nanostructure, randomly distributed nanostructure, metamaterials incorporated OLED, bioinspired nanostructures, etc.\cite{Zhang:21,kim2015enhanced,zhu2016facile,kang2021nanoslot,kim2012biologically}. Out-coupled efficiency (OCE) of OLED represents the ratio of the number of photons extracted outside to the number of recombinations of electrons and holes. In conventional OLED, this efficiency is nearly 20\%\cite{greenham1994angular}. Since EQE of an OLED is the product of its internal quantum efficiency (IQE) and OCE, enhancing the OCE is essential for improving overall device performance. According to Snell’s law, approximately 24\% of the trapped optical modes in bottom-emitting OLEDs result from total internal reflection (TIR) at the interface between the glass substrate and air, caused by their refractive index mismatch\cite{lee2003high}. Random nanoparticle-based composite films can be used as an out-coupling layer for flexible OLED\cite{Shin:20,gu2013light}. Incorporating periodic nanostructures, for example, 2D photonic crystal of \(SiO_{2}/SiN_{x}\) incorporated in glass substrate using Bragg scattering showed significant improvement of up to  in OCE compared to conventional OLED\cite{do2004enhanced}. UV-curable acrylate and sol-gel process, along with planarized ZnO layers, exhibited almost 38\% of OCE improvement in OLED\cite{cho2010solution}. Sung et al. fabricated conical \(SiO_2\) on the substrate of an OLED. After planarization with \(TiO_2\) nanoparticles, 25\% of improvement in OCE was observed\cite{sung2017improved}. In addition to photonic crystals, 2D nanomesh was employed in OLEDs to increase extraction efficiency\cite{ding2014plasmonic}. Moreover, inverse design-based optimization techniques have been used in different photonic devices and applications, such as the design of mode converters, optical switches, photonic crystals, etc.\cite{wei2024inverse, minkov2020inverse}

Here, we report the particle swarm optimization (PSO) based inverse design technique for nanostructures to be incorporated in OLED surface. These periodic nanorods have three layers of metal-insulator-semiconductor oxide (MIS), which significantly improves the LEE of bottom-emitting OLED through the guided surface plasmon coupled emission (SPCE). The major advantage is that LEE enhancement can be wavelength-selective by changing the structural parameters of the nanorod. The semiconductor-oxide layer will reduce TIR by directing optical modes to the metal-insulator interface of the nanorod, which will increase SPCE from the OLED. This approach aims to offer a new strategy for achieving higher optical outcoupling in bottom-emitting OLED devices along with improved emission intensity, spectral luminous efficiency (SLE), UV light suppression, and uniform directionality.

\section{Loss Mechanisms in OLED and Guided SPCE Theory}
\begin{figure}
    \centering
    \includegraphics[width=1\linewidth]{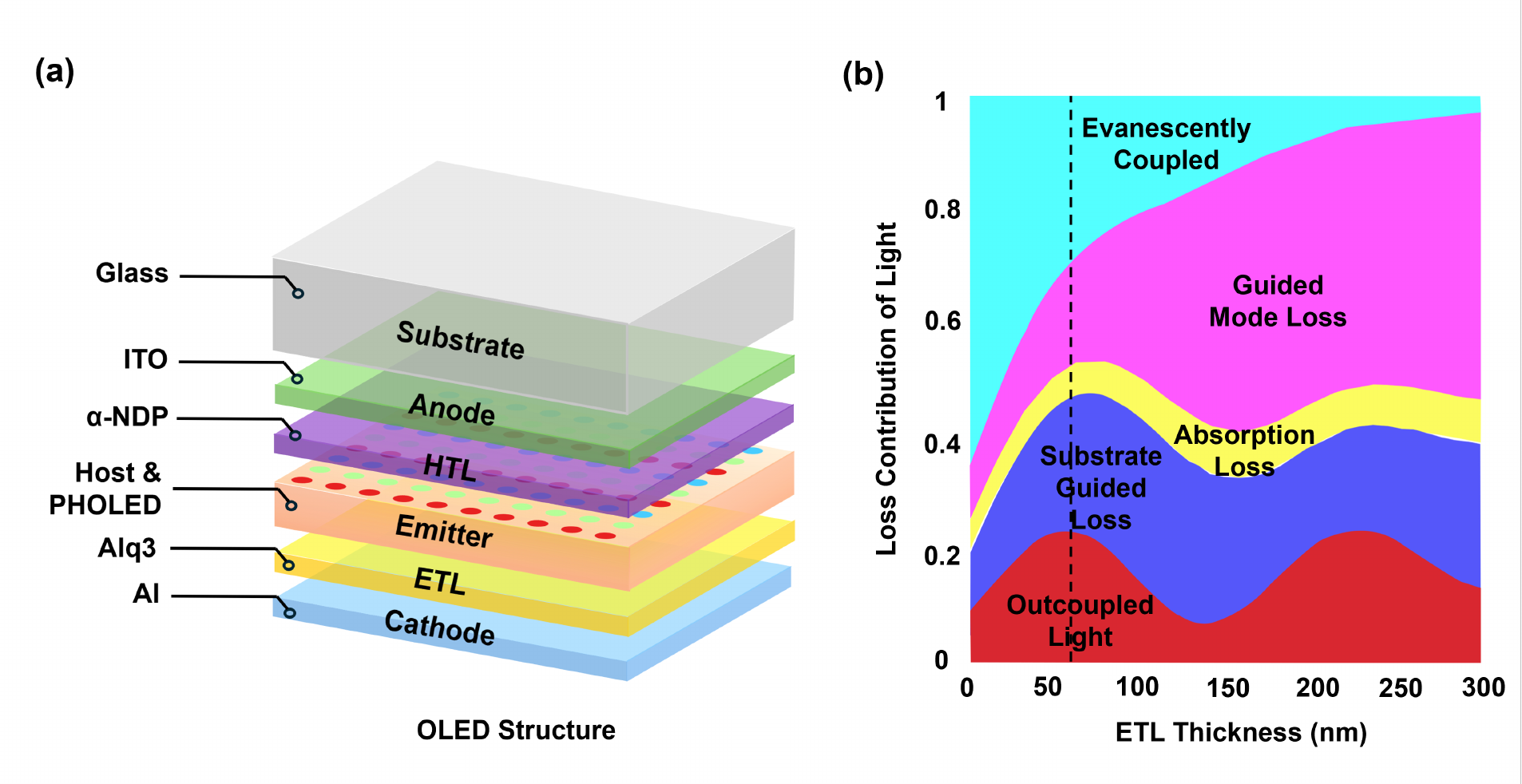}
    \caption{(a) A commercially available OLED structure with different layers (b) Different types of loss contribution inside the conventional OLED. Only 20\% of the light becomes out-coupled to air.}
    \label{fig:1}
\end{figure}

In OLEDs, EQE is achieved as the multiplication of IQE and OCE, \(\eta_{EQE}=\ \eta_{IQE}\times\ \eta_{OCE}\). The OCE decreases due to the reflection loss in the glass substrate, known as substrate-guided loss. For a commercially available OLED structure like the one shown in Fig.~\ref{fig:1}(a), this loss contributes to almost 20\% loss of light inside the OLED structure. Only 20\% of the light is outcoupled to air according to the Fig.~\ref{fig:1}(b). When light is extracted from glass substrate to air, it faces reflection due to the refractive index mismatch between glass \((RI=1.45)\) and air \( (RI=1)\). 
According to Snell's law, if the light incident angle becomes greater than the critical angle of the interface of two media, it faces TIR loss inside the glass medium. That's why nanorod-shaped waveguiding medium like ZnO (Semiconductor Oxide) with higher refractive index can be integrated at the surface of the glass substrate to support more number of optical modes inside the waveguide. The plasmonic nanorod structure, which consists of metal-dielectric interface on top of ZnO layer, can emit the guided optical modes through the creation of surface plasmon, and eventually increase OLED's light extraction\cite{maier2007plasmonics}. 
\begin{figure}[t]
    \centering
    \includegraphics[width=1\linewidth]{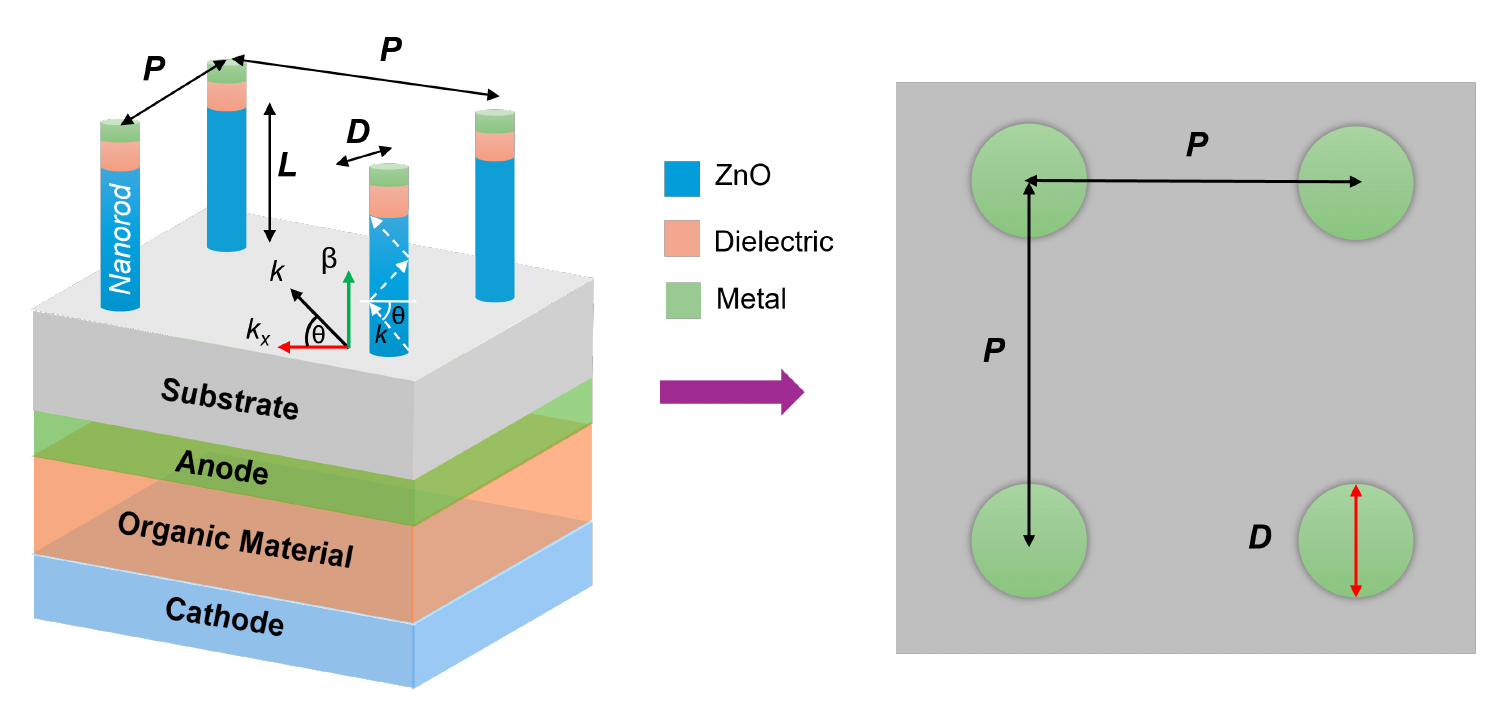}
    \caption{MIS nanorod array placed on the glass substrate. Their structural parameters - ZnO thickness (L), Nanorod diameter (D), Periodicity of array (P) affect the light propagation constants and optical modes inside the nanorod.}
    \label{fig:2}
\end{figure}
In Fig.~\ref{fig:2}, we have a MIS nanorod array of diameter 'D', ZnO thickness 'L', and periodicity 'P'. 'k' represents the wavevector of the mode propagating inside the waveguide. The orthogonal components '$\beta$' and '$k_{x}$' are related to the wavevector through the incident angle '$\theta$', which represents their corresponding mode number. Here, \( k_x = k \cos\theta \; and \; \beta = k \sin\theta \). Since this is a finite cylindrical waveguide, mode trapped inside the waveguide will depend on both axial and transverse boundary conditions\cite{benson2012fields,liao1990microwave}. The transverse and axial resonance condition for the cylindrical waveguide is,
\begin{equation}
k_x = \frac{P_{nm}}{D} \; ; \; \beta = \frac{l\pi}{L}
\label{eq:1}
\end{equation}

Now, from Fig.~\ref{fig:2}, it can be written that

\begin{equation}
\begin{aligned}
& k^2=k_x^2+\beta^2 \\
& \gg\left(\frac{2 \pi}{\lambda}\right)^2=\left(\frac{P_{n m}}{D}\right)^2+\left(\frac{l \pi}{L}\right)^2
\end{aligned}
\label{eq.2}
\end{equation}

Here, $P_{nm}$ refers to the $m-th$ root of the Bessel function of order $n$, which arises from solving the wave equation in cylindrical coordinates with boundary conditions. $l$ is the mode number which can be $l=0,1,2,3,..$ for TM modes and $l=1,2,3,..$ for TE modes. According to Eqn.~\ref{eq.2}, if the length of the waveguide and the number of modes are fixed, to keep the propagation constant $\beta$ real, with increasing wavelength, the peak value of the diameter has to increase. Again, with a fixed wavelength, there is a trade-off between the number of supported modes in the waveguide and the absorption area of the nanorod in the case of a larger diameter. Now, with a fixed diameter, a longer waveguide will be able to support a larger number of modes at a specific wavelength. Experimental results also verify this relation between longer ZnO nanorods and light extraction\cite{Jeong:15}. Furthermore, dipolar coupling for plasmonic resonance exhibits a red shift with a smaller lattice constant; the peak periodicity for light extraction decreases with increasing wavelength\cite{maier2007plasmonics}. 

So, some important performance parameters of OLED, such as radiative decay rate, LEE, angular radiation pattern, color temperature, SLE, and UV light suppression, are affected by the structural parameters of the nanorod array. In this study, we will measure LEE enhancement from the ratio of the farfield transmittance of the MIS nanorod-patterned OLED and the no-patterned OLED structure. Another performance parameter, SLE, can also be measured by the optical power achieved normalized to the human eye sensitivity.

\section{Design Principle}
\subsection{Device Architecture}
\begin{figure}[h]
    \centering
    \includegraphics[width=1\linewidth]{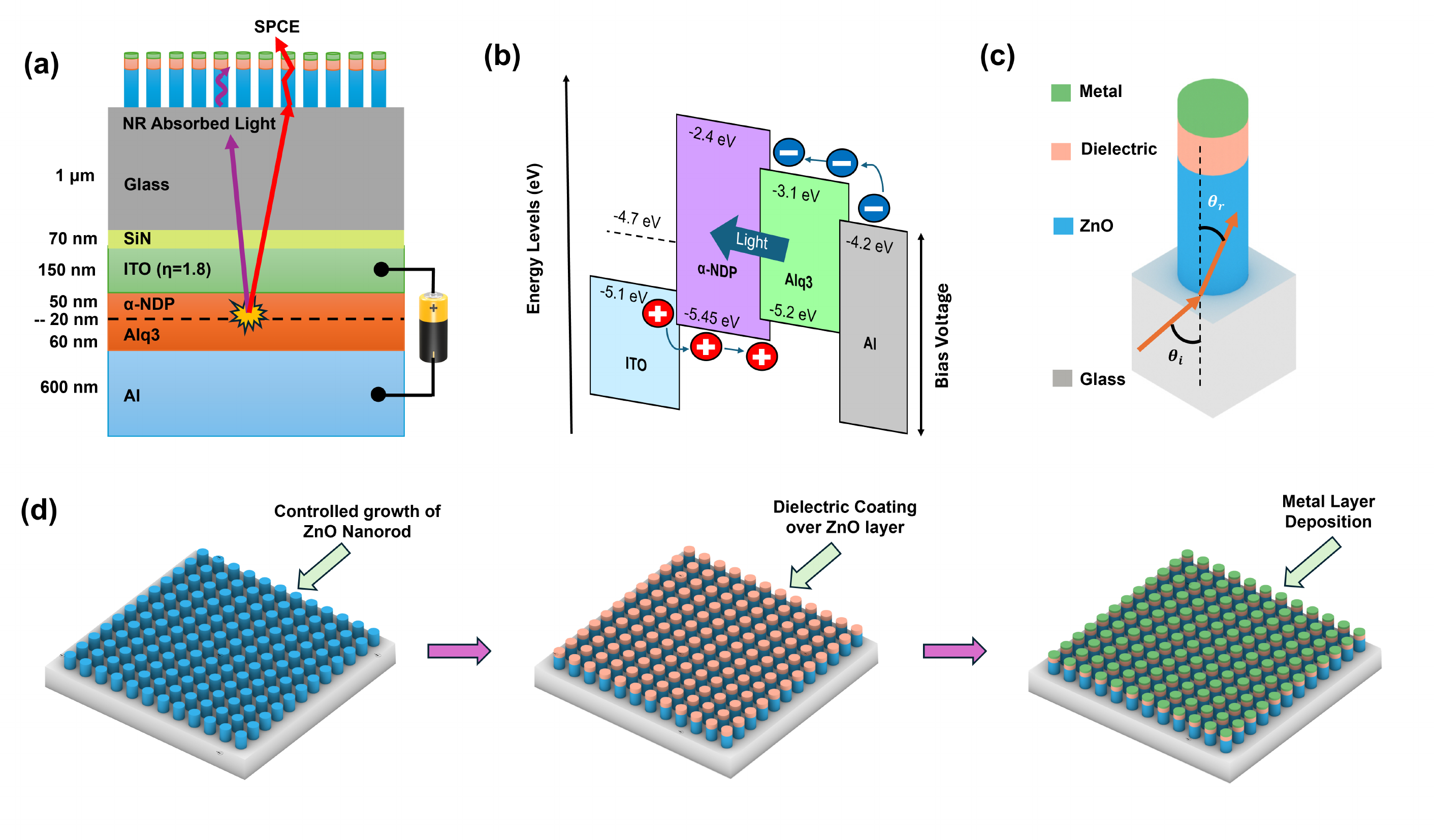}
    \caption{(a) Thickness of different layers in OLED with MIS nanorod placed on the surface of glass substrate which enhances light extraction through SPCE. (b) Energy band diagram of the OLED device. (c) A single nanorod structure made of ZnO, dielectric, and Metal (Ag/Au/Cu) layers. (d) Growth process of the MIS nanorod array on the substrate layer. Controlled growth of ZnO nanorods will be followed by dielectric coating. Finally, the metal layer will be deposited.}
    \label{fig:3}
\end{figure}

An OLED is an injection-type device comprising two electrodes separated by an organic multilayer. The thickness of different layers and the energy band diagram of our studied OLED device are illustrated in Fig.~\ref{fig:3}(a)-(b). The used OLED device employs Al as the cathode, Alq$_3$ as the electron transport layer (ETL), $\alpha$-NPD as the hole transport layer (HTL), ITO as the anode, SiN as the encapsulation layer, and glass as the substrate. For evaluating LEE enhancement, identical OLED structures were simulated with and without an MIS nanorod array on the substrate. A single MIS nanorod structure and the growth process of the nanorods are illustrated in Fig.~\ref{fig:3}(c)-(d). 

ZnO was selected as the semiconductor oxide, while the dielectric and metal thickness and dielectric material were determined from broadband transmittance across the visible range by changing those parameter values.

\subsection{Optimization Algorithm}
\begin{figure}[t]
    \centering
    \includegraphics[width=0.8\linewidth]{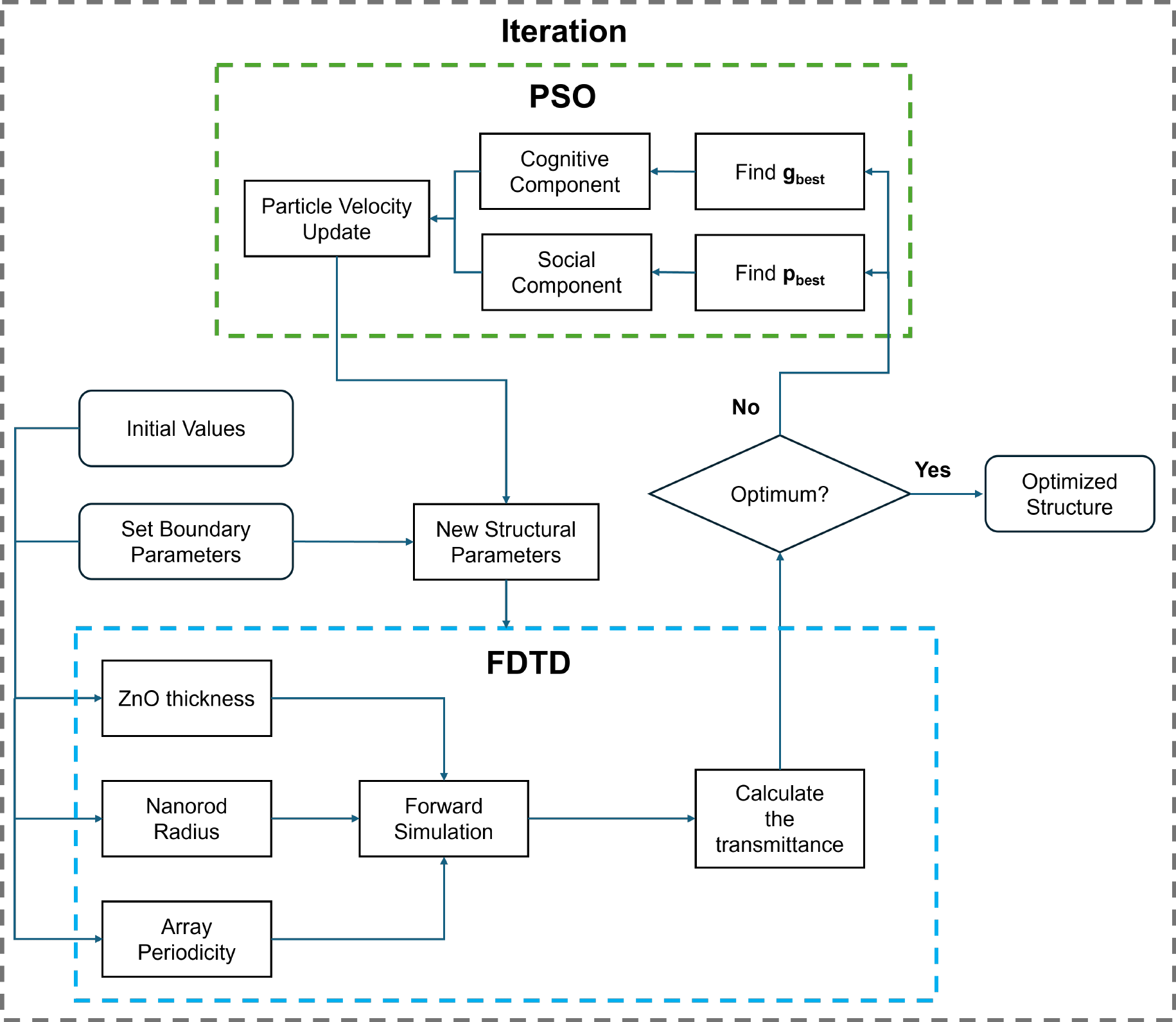}
    \caption{The algorithm flow chart of inverse design optimization of OLED with MIS nanorods, which provides optimal parameter values of the structure.}
    \label{fig:4}
\end{figure}

The structural parameters of the nanorod array (ZnO thickness, Nanorod radius, and Periodicity) have been optimized using inverse-design optimization technique at three distinct wavelengths of 440 nm, 550 nm, and 640 nm for blue, green, and red emitting OLEDs. To observe the trend of LEE enhancement compared to conventional OLED for varying a single parameter, every parameter is varied in a specific range. Then the boundary values for these parameters are set for the optimization algorithm, which is known as the Particle Swarm Optimization (PSO) algorithm. The algorithm flow chart to achieve the optimum parameter values is depicted in Fig.~\ref{fig:4}. The algorithm consists of one judgmental unit (optimum) and two processing units (PSO and FDTD). In the PSO unit, every particle will find its personal best parameters and compare them with the global parameters. Then the cognitive and social components will be added to the current velocity of the particles, which is multiplied by the inertia weight w. Thus, the algorithm will adjust the new position of the particles (new structural parameters).

\begin{equation}
\begin{gathered}
v_i(t+1)=w v_i(t)+c_1 r_1\left(p B e s t_i-x_i\right)+c_2 r_2\left(g B e s t-x_i\right) \\
x_i(t+1)=x_i(t)+v_i(t+1)
\end{gathered}
\label{eq.5}
\end{equation}
Here, r1 and r2 are random matrices regenerated after every iteration and is bounded in $[0,1]$ range, and c1 and c2 are coefficients for cognitive and social components. Based on the requirement of increasing LEE, the near-field transmission of light, $T_{near-field}$,  which is the figure of merit in the optimization algorithm, is monitored by a field monitor placed above the OLED structure, completing forward simulations in FDTD. The figure of merit is defined as follows:
\begin{equation}
\text { maximize } f(L, D, P)= |T_{near-field}|
\label{eq.6}
\end{equation}

The Finite Difference Time Domain (FDTD) method was employed to solve Vector 3D Maxwell's equations for the forward simulations with the parameter values obtained from optimization. The Perfectly Matched Layers (PML) boundary conditions were applied to reduce the computational load and avoid the unnecessary reflection of light at the computational boundaries, and metal boundary conditions were used in the cathode layer. Multiple dipole sources were positioned using unit cell symmetry to simulate the behaviour of light inside the OLED structure\cite{lu2016plasmon}. Different locations and orientations of dipole sources have been implemented through parameter sweep. A frequency domain field profile monitor captured the data of time domain electric and magnetic fields and the transmitted power near the structure. Then the time-domain field was converted to the frequency domain using the Fourier transform. These frequency domain fields are projected into the far-field through directional vectors accounting for Snell's law and Fresnel's law of reflection occurring at the far-field interface.

\section{Results and Analysis}
\subsection{Guided SPCE}
The effect of metal layer thickness in the nanorod on SPP generation was first investigated, given its direct impact on light absorption. In addition, variations in the dielectric refractive index and thickness were explored to find the optimum plasmonic structure.
\begin{figure}[h]
    \centering
    \includegraphics[width=1\linewidth]{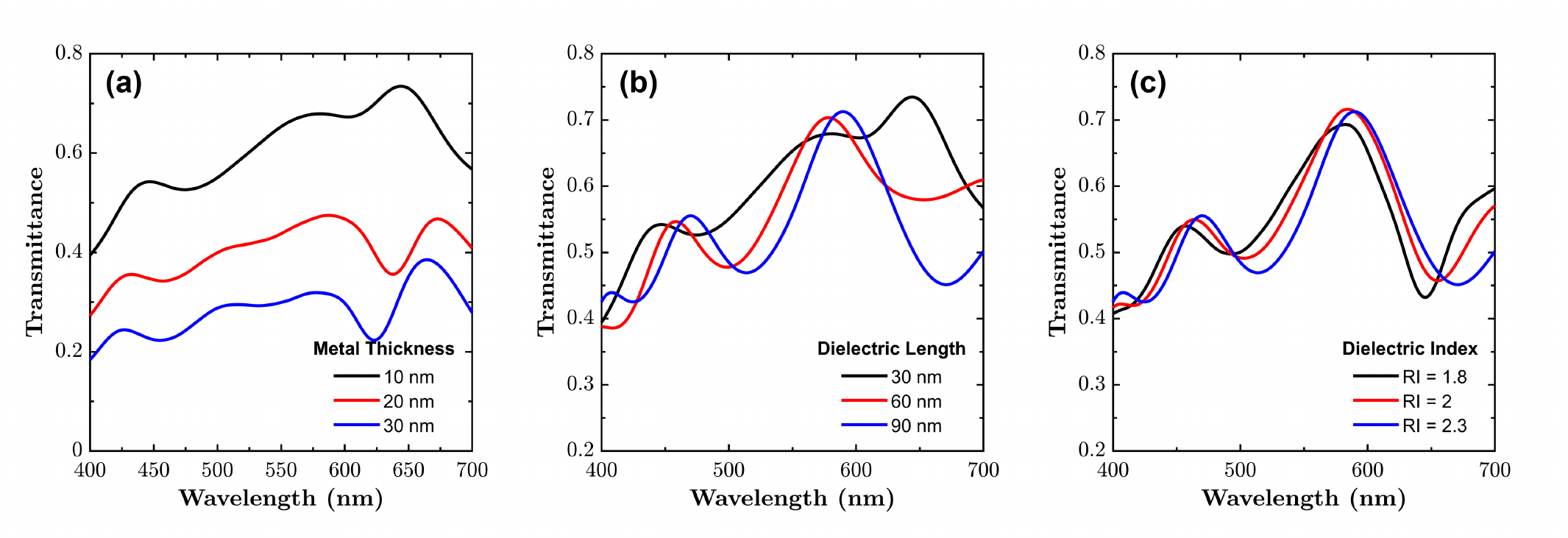}
    \caption{Transmission spectra of the MIS nanorod structure under varying meta-dielectric design parameters: (a) Metal layer thickness (10 nm, 20 nm, 30 nm); (b) Dielectric layer thickness (30 nm, 60 nm, 90 nm); and (c) Refractive index of the dielectric layer (RI = 1.8, 2.0, 2.3). }
    \label{fig:5}
\end{figure}
From Fig.~\ref{fig:5}(a), it is seen that 10 \(nm\) metal thickness results in maximum light transmission in the visible range. 
\begin{figure}[h]
    \centering
    \includegraphics[width=1\linewidth]{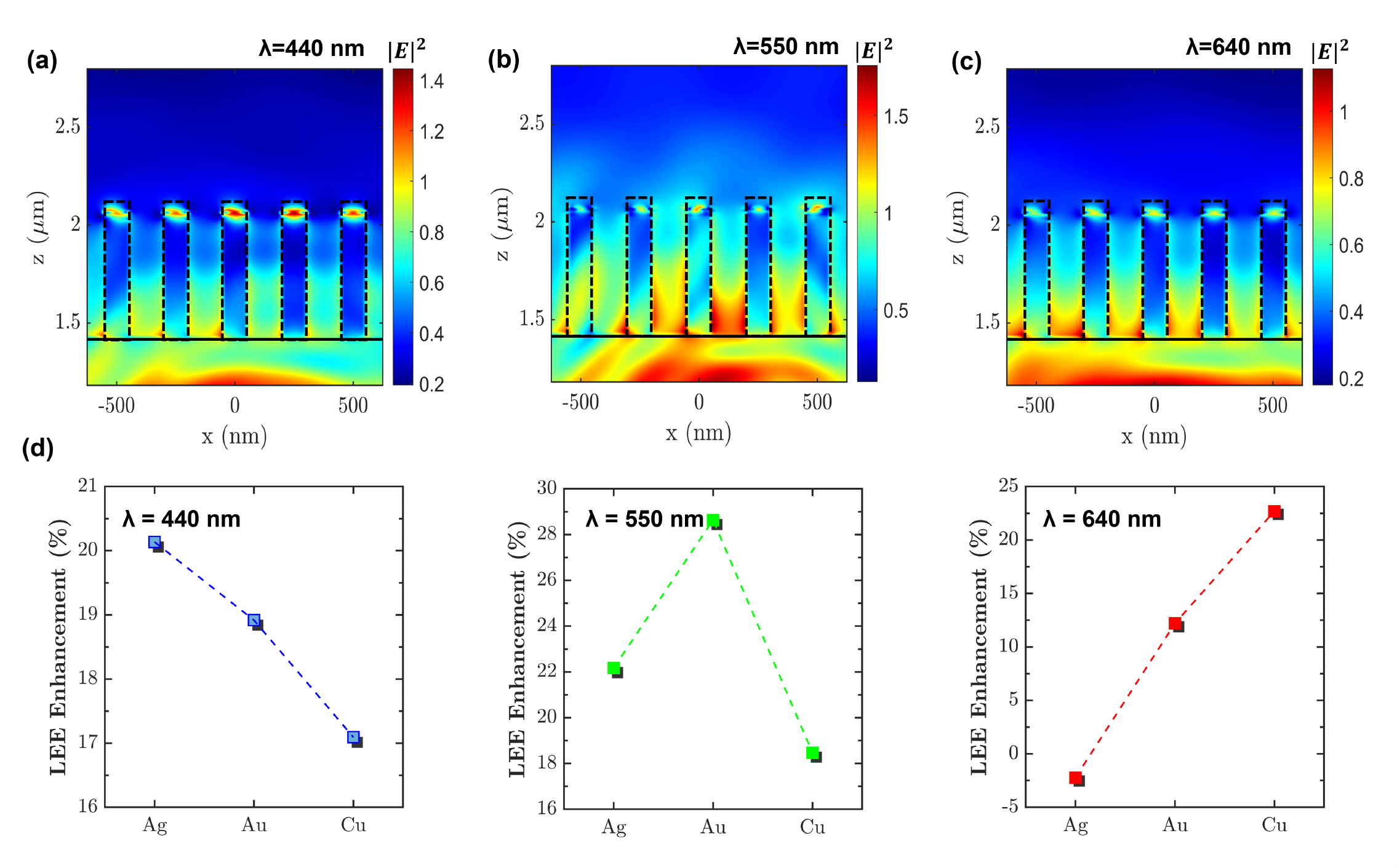}
    \caption{Generation of SPPS at the metal–dielectric interface of the nanorod for different metals from the guided light in ZnO layer: (a) Ag at 440 \(nm\), (b) Au at 550 \(nm\), and (c) Cu at 640 \(nm\). (d) LEE enhancement trend at 440 nm, 550 nm, and 640 nm wavelength for different metals}
    \label{fig:6}
\end{figure}
A steady increase in light transmittance is observed at a dielectric thickness of 30 \(nm\) in Fig.~\ref{fig:5}(b), and the change of refractive index of the dielectric material shows a minimal variation in light transmission in Fig.~\ref{fig:5}(c). From the observations, a metal layer thickness of 10 nm, $Al_{2}0_{3}$, RI = 1.7$\sim$1.8 in visible range, as dielectric layer, and dielectric thickness of 30 nm have been used as the plasmonic section of the structure for optimization process. 

To choose the metal, enhancement of light intensity at the metal-dielectric interface for different metals (Au, Cu, Ag) is observed in Fig.~\ref{fig:6}(a-c), which confirms SPP generation from guided light wave. Ag exhibits strong surface plasmon confinement in the wavelength range of 400--450\,nm, while Au demonstrates maximum confinement near 530\,nm and Cu around 600\,nm. That's why Ag, Au, and Cu are selected for blue, green, and red emission regions, respectively, due to their distinct plasmonic responses at corresponding wavelengths\cite{maier2007plasmonics}. ZnO nanorod as resonator-type waveguide scatters the field to localized light\cite{mangalgiri2017dielectric,van2013designing}. This resonance induced a stronger electric field along the nanorod gaps due to the refractive index contrast between ZnO and air, causing light refraction at the nanorod–air interface per Snell’s law\cite{Hsiao2014,Kim2005}. LEE enhancement also maximizes at 440 nm, 550 nm, and 640 nm wavelengths for the corresponding usage of Ag, Au, and Cu as metal layer, as expected, which is portrayed in Fig.~\ref{fig:6}(d).

\subsection{LEE Enhancement Trend}
After defining the plasmonic structure, the enhancement trend of light extraction efficiency (LEE) at 440, 550, and 640\,nm was investigated by systematically varying the structural parameters of the MIS nanorods, including ZnO thickness (L), nanorod radius (R), and array periodicity (P).

\begin{figure}[h]
    \centering
    \includegraphics[width=0.85\linewidth]{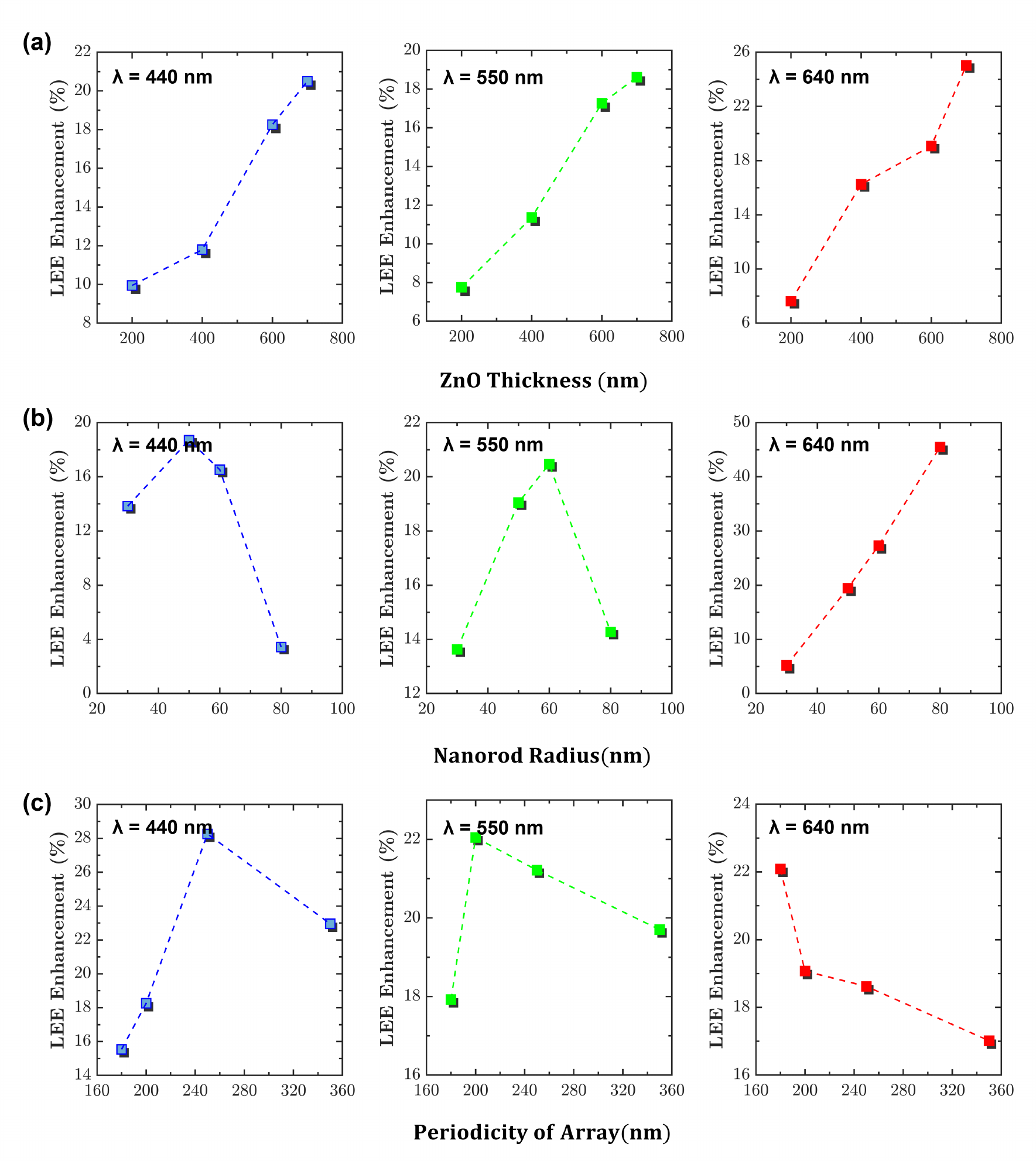}
    \caption{LEE enhancement trend at 440 nm, 550 nm, and 640 nm wavelength for various structure parameters: (a)ZnO thickness (b)Nanorod radius (c)Periodicity of Array}
    \label{fig:7}
\end{figure}

As shown in Fig.~\ref{fig:7}(a), the LEE enhancement curve increases with ZnO thickness for the three wavelengths. In Fig.~\ref{fig:7}(b), the optimal nanorod radius shifts to larger values with increasing wavelength (red shift). However, at a fixed wavelength, the peak radius decreases beyond a critical point because of the larger absorption area for the photon. 
Finally, Fig.~\ref{fig:7}(c) demonstrates that the peak periodicity shifts to a smaller value as the wavelength faces a red shift. The enhancement trends for different structural parameters with respect to wavelength shifting follow the theoretical estimation depicted in the previous section with the Eqn.~\ref{eq.2}.

\subsection{Optimization Results}
\begin{figure}[h]
    \centering
    \includegraphics[width=1\linewidth]{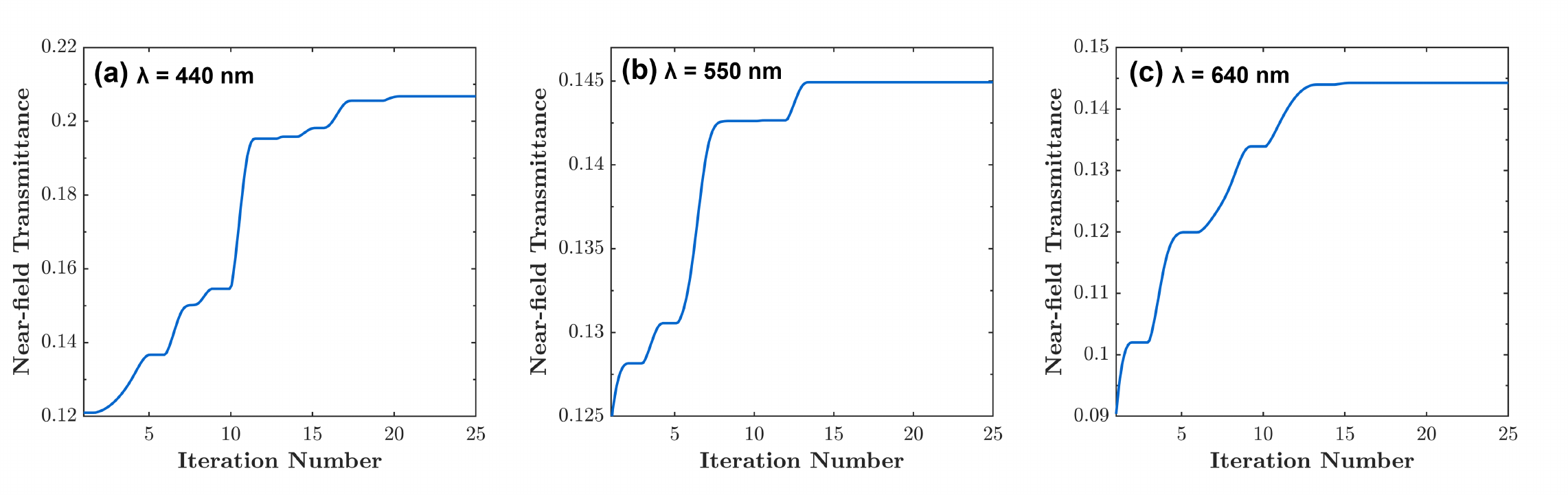}
    \caption{Updated value of near-field transmittance  after each iteration from PSO algorithm at (a) 440 nm (b) 550 nm (c) 640 nm wavelength}
    \label{fig:8}
\end{figure}

Observing the parameter trends, the boundary values set for the PSO algorithm are: ZnO thickness [100 nm $\sim$ 800 nm], Nanorod radius [30 nm $\sim$ 80 nm], Array Periodicity [180 nm $\sim$ 350 nm]. After 25 iterations, the objective function reaches a converged value and provides optimized structure parameters for all three wavelengths, as shown in Fig.~\ref{fig:8}. 

\begin{table}[h!]
\centering
\resizebox{0.95\textwidth}{!}{
\renewcommand{\arraystretch}{1.4} 
\begin{tabular}{ccccccc}
\hline
\textbf{Wavelength} 
& \makecell{\textbf{ZnO}\\\textbf{Thickness}\\(nm)} 
& \makecell{\textbf{Nanorod}\\\textbf{Radius}\\(nm)} 
& \makecell{\textbf{Array}\\\textbf{Periodicity}\\(nm)} 
& \textbf{Metal} 
& \makecell{\textbf{Max}\\\textbf{Transmittance}} 
& \makecell{\textbf{LEE}\\\textbf{Enhancement}\\(\%)} \\
\hline

\multirow{2}{*}{\(\lambda = 440 \,\text{nm}\)} 
& 150.31 & 74.00 & 250.00 & \multirow{2}{*}{Ag} & \multirow{2}{*}{0.2066} & \multirow{2}{*}{29.12} \\
& 100.00 & 80.00 & 350.00 &                     &                       &                       \\
\hline

\multirow{2}{*}{\(\lambda = 550 \,\text{nm}\)} 
& 105.67 & 78.89 & 204.06 & \multirow{2}{*}{Au} & \multirow{2}{*}{0.1447} & \multirow{2}{*}{30.36} \\
& 100.00 & 80.00 & 180.00 &                     &                       &                       \\
\hline

\multirow{2}{*}{\(\lambda = 640 \,\text{nm}\)} 
& 136.48 & 68.30 & 195.03 & \multirow{2}{*}{Cu} & \multirow{2}{*}{0.1442} & \multirow{2}{*}{51.78} \\
& 120.26 & 74.97 & 185.75 &                     &                       &                       \\
\hline
\end{tabular}
}
\caption{Optimized structural parameters of MIS nanorod array for different metals}
\label{tab:1}
\end{table}
Since multiple parameters affect the transmittance, there can be more than one optimized combination to achieve the optimal figure of merit. The combinations of optimized parameters to achieve the maximum objective function are listed in Table.~\ref{tab:1}. If we look at the optimized parameters, for example, at 440 nm wavelength, ZnO thickness has reduced while the nanorod radius has increased in the 2nd combination compared to the 1st one. This trend has also been followed for the combinations in 550 and 640 nm wavelengths. So, the optimized parameter values reconfirm the Eqn.~\ref{eq.2} that for resonance at a wavelength, if one parameter between $L$ and $D$ goes to the maximum value, the other one will try to reach the minimum to incorporate maximum optical mode. By reducing the nanorod length and increasing their radius, the proposed structure can enhance light extraction, decreasing the vertical footprint of the device relative to previously reported results at the same time.\cite{kwon2018plasmonic}

With the optimized MIS nanorod structure on the OLED surface, other key performance parameters, including emission intensity, color temperature, far-field angular distribution, SLE, transmittance in UV wavelength, and electrical characteristics, were also analyzed. 

\subsection{Optical Performance}
\begin{figure}[h]
    \centering
    \includegraphics[width=1\linewidth]{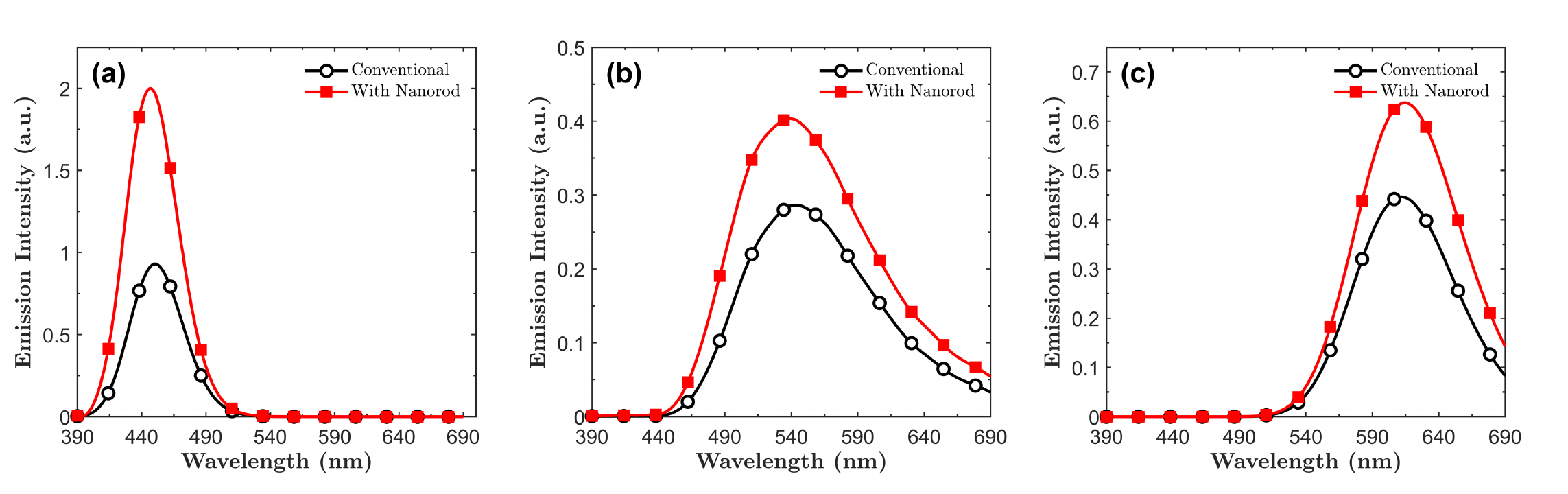}
    \caption{Enhanced emission intensity after incorporating MIS nanorod on the glass surface for (a) blue (b) green (c) red emitting OLED devices.}
    \label{fig:9}
\end{figure}

\begin{figure}[h]
    \centering
    \includegraphics[width=1\linewidth]{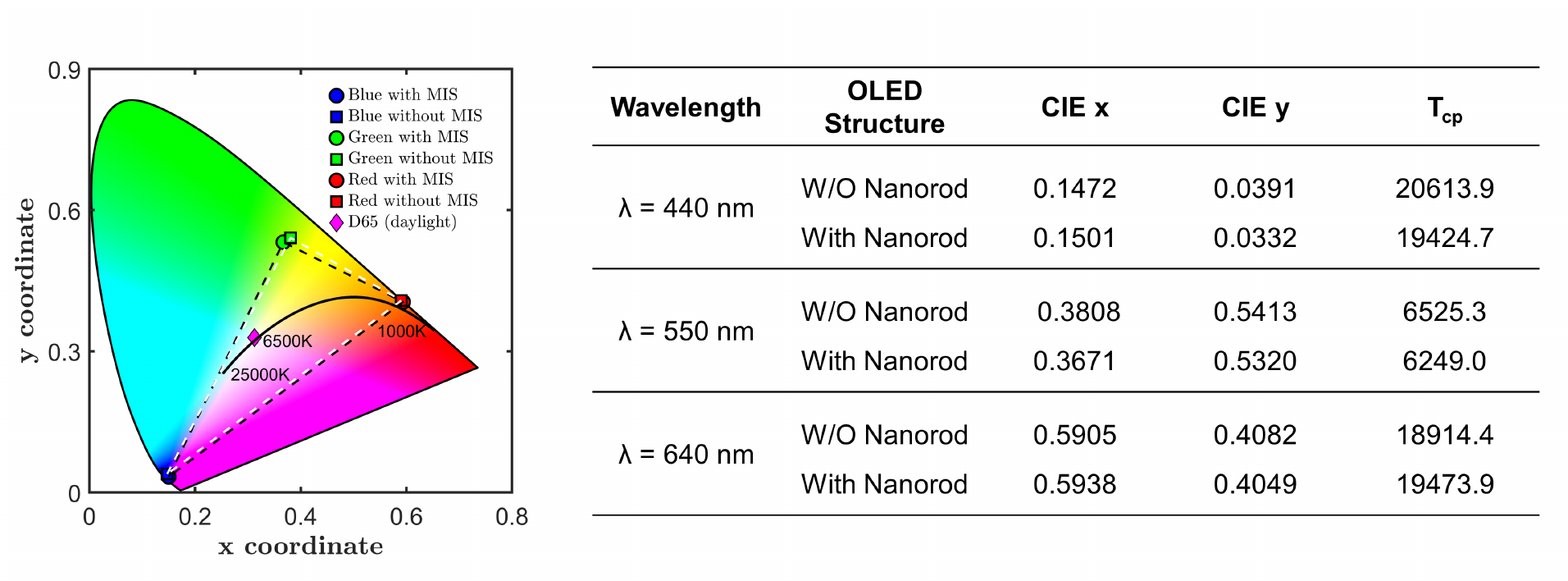}
    \caption{CIE coordinates of the resultant OLED device with MIS nanorod on the substrate surface and without the nanorod array. All corresponding values are listed in a table, relative to those of the reference sample without
ZnO nanorods.}
    \label{fig:10}
\end{figure}

As shown in Fig.~\ref{fig:9}(a-c), the optimized MIS nanorods enhance emission intensity across the blue, green, and red spectral regions. Maximum emission intensity increases for red OLED by almost 2 times. It also increases by 1.33 times and 1.42 times for green and blue OLED, respectively. The linewidth of the spectrum has become broader for all the OLEDs. 
In the case of OLED color quality,as shown in Fig~\ref{fig:10}, blue and red OLEDs are in deep saturation regions. So, their CCT values don't lie on the Planckian locus range. 
But green OLED gives CCT values of 6249 K with MIS nanorod array and 6525.3 K for OLED without nanorod. These values are close to the broad daylight white color (D65), which indicates the good color quality of the green OLED for display applications. The color triangle does not change much for the OLED with the MIS nanorod array compared to the conventional one.
\begin{figure}[h]
    \centering
    \includegraphics[width=1\linewidth]{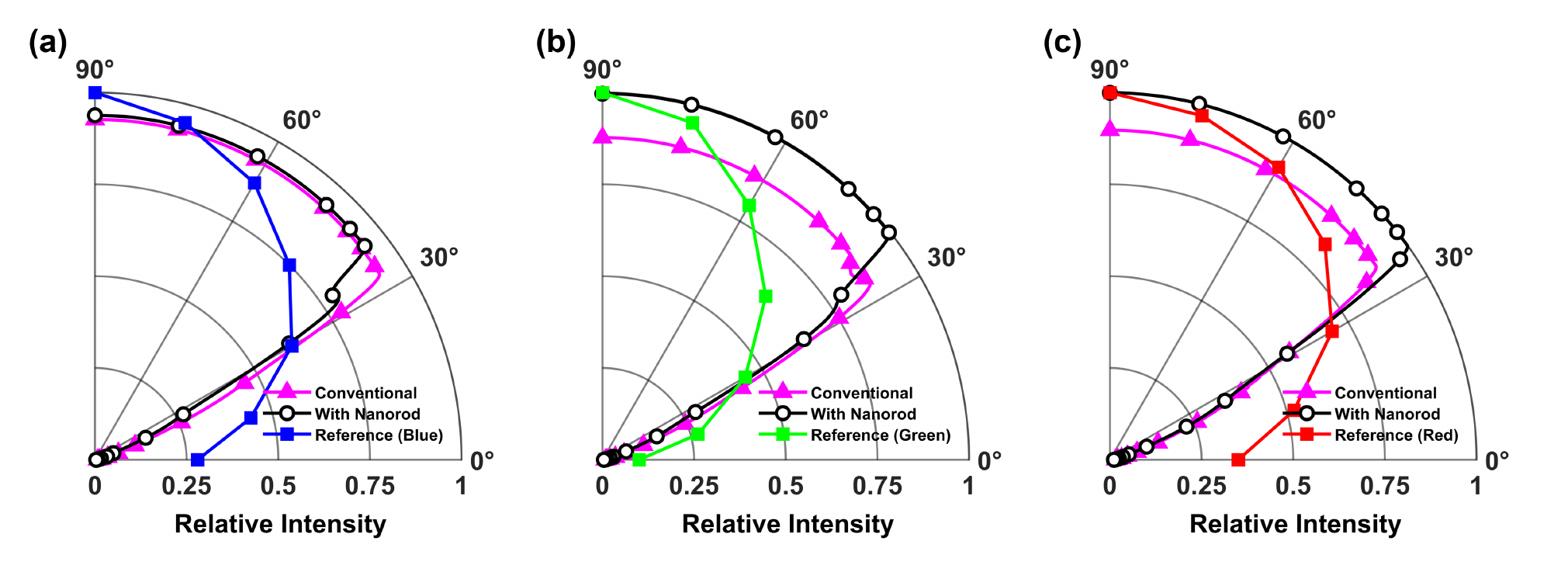}
    \caption{Comparative analysis of angular radiation pattern between conventional OLED, MIS nanorod integrated OLED, and a refernce OLED at (a) \(\lambda\) = 440 \(nm\), (b) \(\lambda\) = 550 \(nm\), (c) \(\lambda\) = 640 \(nm\) wavelengths}
    \label{fig:11}
\end{figure}

Subsequently, the angular radiation intensity was analyzed for the simulated OLED without a nanorod array, with the optimized nanorod array, and for a reference OLED reported in the literature\cite{https://doi.org/10.1002/sdtp.13343}. The angular coverage of the light intensity of these three devices at wavelengths of 440 nm, 550 nm, and 640 nm is portrayed in Fig.~\ref{fig:11}(a-c). At $\lambda = 440$ nm, the radiation pattern shows no change after incorporating the MIS nanorod array, whereas at $\lambda = 550$ nm and $\lambda = 640$ nm the incorporation of the MIS nanorods leads to a more intense light distribution keeping the uniformity from 35$^{\circ}$ to 90$^{\circ}$ angles. The effective viewing angle for the proposed device expands up to 60$^{\circ}$.

Another performance parameter, the spectral luminosity efficiency (SLE), was measured at 440, 550, 640 nm wavelengths for the optimized MIS nanorod structures with incorporated OLED surface, demonstrated in Fig.~\ref{fig:10}(a). The SLE values obtained at emission wavelengths of 440~nm, 550~nm, and 640~nm are 5.07\%, 62.57\%, and 45.57\%, respectively. 
\begin{figure}[h]
    \centering
    \includegraphics[width=0.8\linewidth]{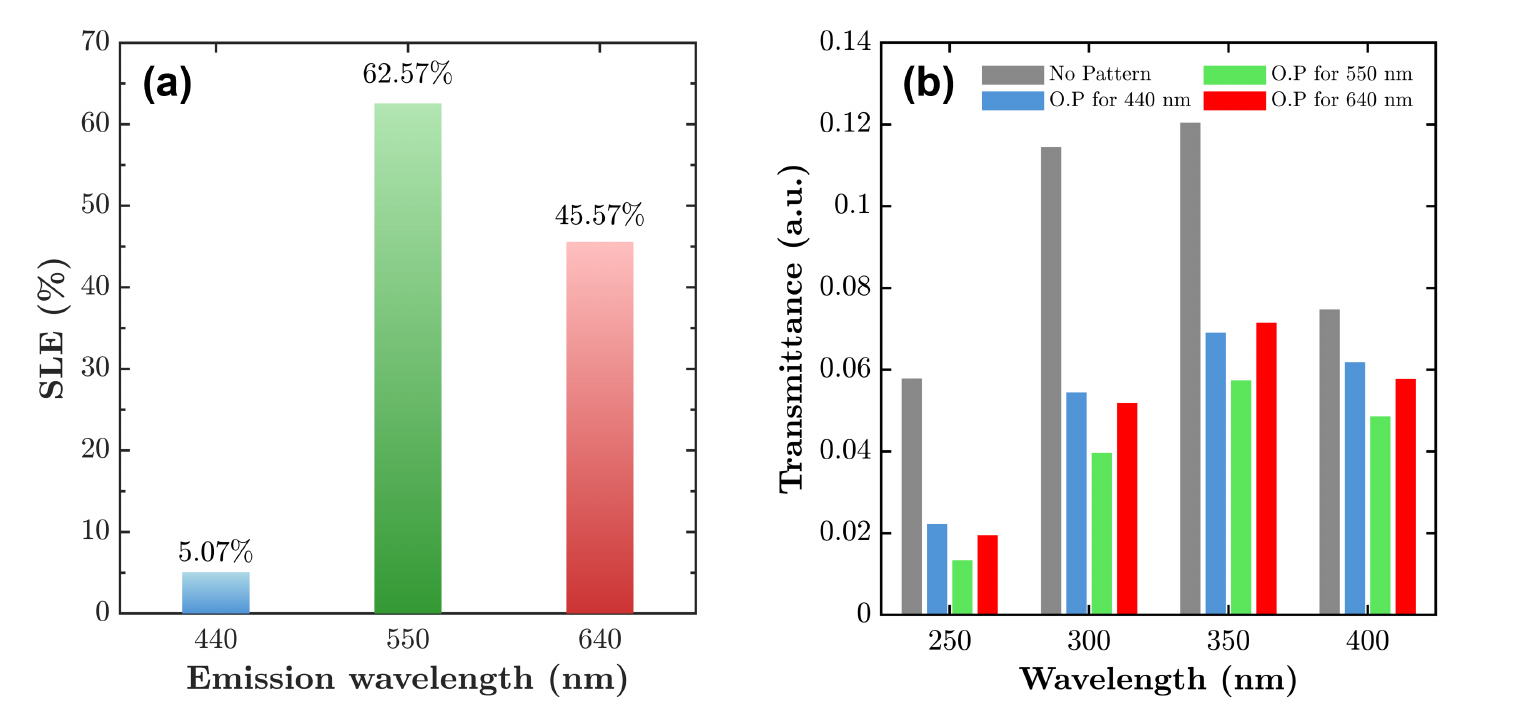}
    \caption{(a) SLE values at 440~nm, 550~nm, and 640~nm emission wavelengths for optimized MIS nanorod structures incorporated on the OLED surface. (b) UV transmittance reduces at 250~nm, 300~nm, 350~nm, and 400~nm wavelengths by using MIS nanorods, which were optimized for blue (440~nm), green (550~nm), and red (640~nm) emission, compared to the conventional OLED structure without nanorods.}
    \label{fig:12}
\end{figure}
The highest SLE is obtained at a wavelength of 550~nm, which corresponds to the maximum overlap with the human eye's luminosity function. In contrast, the blue emission at 440~nm exhibits the lowest SLE, primarily due to fluorescence-related limitations\cite{nano13182521}.

Another advantage of incorporating optimized nanorods on the glass surface is the effective suppression of UV wavelengths. The optimized parameters corresponding to emission wavelengths of 440~nm, 550~nm, and 640~nm significantly decrease the UV light output from OLEDs compared to conventional structures. The transmittance in this region reduces from 12\% to 5\% - 7\% because of the nanorod structure on top. In particular, the optimized MIS nanorod design for 550~nm emission reduces the transmittance to below 5\% in the deep UV region, representing a comparable improvement with the state-of-the-art~\cite{Chen:25}. The reduction of UV light at 250~nm, 300~nm, 350~nm, and 400~nm wavelengths achieved by introducing optimized nanorods for blue, green, and red emission on the OLED surface is presented in a grouped bar chart in Fig.~\ref{fig:10}(b).

\subsection{Electrical Performance}
We have also compared the electrical characteristics and found no change in internal quantum efficiency (IQE) and J-V plot between conventional OLED and OLED with MIS nanorod, as shown in Fig.~\ref{fig:12}.
\begin{figure}[h]
    \centering
    \includegraphics[width=0.8\linewidth]{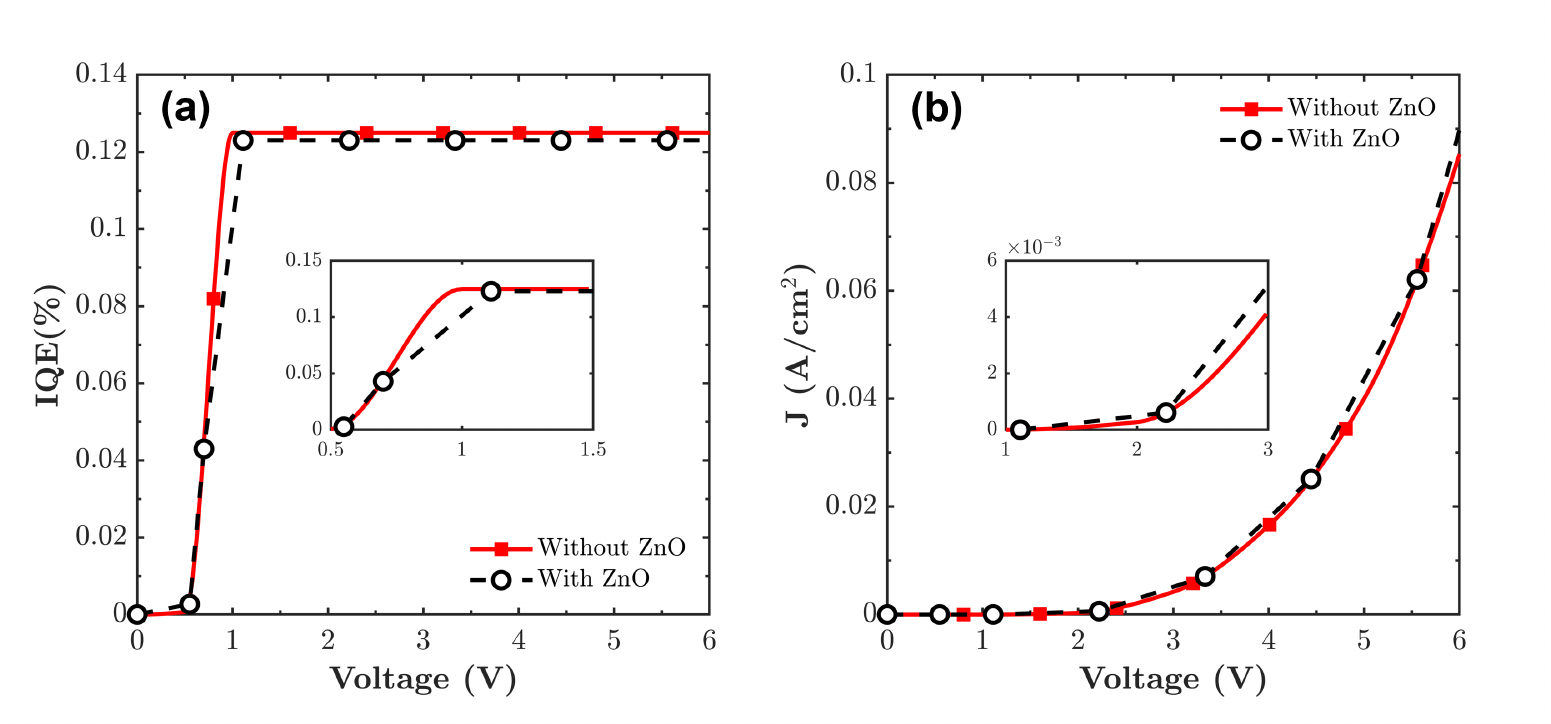}
    \caption{Comparative analysis of (a) IQE (b) J-V characteristics between OLED with and without MIS nanorod. The voltage at which IQE becomes saturated and the cut-off voltage are portrayed in the inset of the graphs}
    \label{fig:13}
\end{figure}
Since the MIS nanorod array on the glass surface is a light extraction structure, it will not have any impact on the photon generation from electron-hole pair recombination in the emitting layer, and the photon recycling effect is also not considered. From the inset of the graphs, the maximum IQE for our studied OLED structure is 12.5\% and the cutoff voltage is found to be 1.8 V.

\section{Fabrication Feasibility}
For the growth of ZnO nanostructures, various techniques are used, including pulsed laser deposition\cite{sun2007production,taschuk2008angularly},chemical vapor deposition\cite{thandavan2015enhanced}, electrochemical deposition\cite{sun2013morphology}, hydrothermal\cite{dong2012fluorescent}, atomic layer deposition\cite{nandanapalli2018surface} and chemical bath deposition\cite{zheng2024electrically}. Compared with many of the previously mentioned techniques, chemical bath deposition (CBD) is a relatively low-temperature and low-cost ZnO synthesis method based on a relatively simple procedure. Deposition of metal and dielectric layers can also be achieved through methods like e-beam evaporation and sputtering \cite{hackett2018spectrometer}

\section{Conclusion}
Integration of optimized periodic MIS nanorod arrays on the substrate surface of OLED substrate has led to a significant enhancement in LEE through the guided surface plasmon-coupled emission. These findings not only address the total internal reflection losses in the substrate layer of OLEDs but also demonstrate the process for wavelength-selective emission tuning. The LEE enhancement achieved at 440, 550, and 640 nm with increased emission intensity suggests promising applications in high-performance displays and energy-efficient lighting. This inverse design optimization technique can contribute to the future development of nanostructured optoelectronic devices. Future research may explore alternative materials and scalable fabrication for further performance and cost optimization.

\begin{acknowledgement}

The authors acknowledge the financial grant (Basic Research Grant, BUET, Office Order no.: Shongstha/R-60/Re-2413, Dated: 10 Oct 2023 (Professor Dr. Md. Zahurul Islam)) and logistical support provided by the Department of EEE, BUET throughout the duration of this work.

\end{acknowledgement}

\section{Disclosure}

The authors declare no conflicts of interest.

\section{Data Availability}

Data for this article, including LEE enhancement, optimization data are available at $https://github.com/kh-zahin/MIS-Nanorod-OLED-LEE.git$



\bibliography{achemso-demo}

\begin{mcitethebibliography}{42}
\providecommand*\natexlab[1]{#1}
\providecommand*\mciteSetBstSublistMode[1]{}
\providecommand*\mciteSetBstMaxWidthForm[2]{}
\providecommand*\mciteBstWouldAddEndPuncttrue
  {\def\EndOfBibitem{\unskip.}}
\providecommand*\mciteBstWouldAddEndPunctfalse
  {\let\EndOfBibitem\relax}
\providecommand*\mciteSetBstMidEndSepPunct[3]{}
\providecommand*\mciteSetBstSublistLabelBeginEnd[3]{}
\providecommand*\EndOfBibitem{}
\mciteSetBstSublistMode{f}
\mciteSetBstMaxWidthForm{subitem}{(\alph{mcitesubitemcount})}
\mciteSetBstSublistLabelBeginEnd
  {\mcitemaxwidthsubitemform\space}
  {\relax}
  {\relax}

\bibitem[Lin \latin{et~al.}(2022)Lin, Li, Chen, Hsu, Wei, Lee, and Chiu]{Lin:22}
Lin,~B.-Y.; Li,~Y.-R.; Chen,~C.-H.; Hsu,~H.-C.; Wei,~M.-K.; Lee,~J.-H.; Chiu,~T.-L. Effects of electron transport layer thickness on light extraction in corrugated OLEDs. \emph{Opt. Express} \textbf{2022}, \emph{30}, 18066--18078\relax
\mciteBstWouldAddEndPuncttrue
\mciteSetBstMidEndSepPunct{\mcitedefaultmidpunct}
{\mcitedefaultendpunct}{\mcitedefaultseppunct}\relax
\EndOfBibitem
\bibitem[Ulla \latin{et~al.}(2024)Ulla, Kiran, Ansari, Alam, Girma, and Gedda]{ULLA2024114602}
Ulla,~H.; Kiran,~M.~R.; Ansari,~S.~N.; Alam,~M.~W.; Girma,~W.~M.; Gedda,~G. Effect of hole-transport layer thickness on the performance of organic light-emitting diodes. \emph{Optical Materials} \textbf{2024}, \emph{147}, 114602\relax
\mciteBstWouldAddEndPuncttrue
\mciteSetBstMidEndSepPunct{\mcitedefaultmidpunct}
{\mcitedefaultendpunct}{\mcitedefaultseppunct}\relax
\EndOfBibitem
\bibitem[Chen \latin{et~al.}(2010)Chen, Xie, Yang, Chen, and Huang]{chen2010high}
Chen,~S.; Xie,~J.; Yang,~Y.; Chen,~C.; Huang,~W. High-contrast top-emitting organic light-emitting diodes with a Ni/ZnS/CuPc/Ni contrast-enhancing stack and a ZnS anti-reflection layer. \emph{Journal of Physics D: Applied Physics} \textbf{2010}, \emph{43}, 365101\relax
\mciteBstWouldAddEndPuncttrue
\mciteSetBstMidEndSepPunct{\mcitedefaultmidpunct}
{\mcitedefaultendpunct}{\mcitedefaultseppunct}\relax
\EndOfBibitem
\bibitem[Sun and Forrest(2008)Sun, and Forrest]{sun2008enhanced}
Sun,~Y.; Forrest,~S.~R. Enhanced light out-coupling of organic light-emitting devices using embedded low-index grids. \emph{Nature photonics} \textbf{2008}, \emph{2}, 483--487\relax
\mciteBstWouldAddEndPuncttrue
\mciteSetBstMidEndSepPunct{\mcitedefaultmidpunct}
{\mcitedefaultendpunct}{\mcitedefaultseppunct}\relax
\EndOfBibitem
\bibitem[Kova\v{c}i\v{c} \latin{et~al.}(2021)Kova\v{c}i\v{c}, Samigullina, Bouchard, Kr\v{c}, Lipov\v{s}ek, Soldera, Lasagni, Reineke, and Topi\v{c}]{Kovacic:21}
Kova\v{c}i\v{c},~M.; Samigullina,~D.; Bouchard,~F.; Kr\v{c},~J.; Lipov\v{s}ek,~B.; Soldera,~M.; Lasagni,~A.~F.; Reineke,~S.; Topi\v{c},~M. Analysis and optimization of light outcoupling in OLEDs with external hierarchical textures. \emph{Opt. Express} \textbf{2021}, \emph{29}, 23701--23716\relax
\mciteBstWouldAddEndPuncttrue
\mciteSetBstMidEndSepPunct{\mcitedefaultmidpunct}
{\mcitedefaultendpunct}{\mcitedefaultseppunct}\relax
\EndOfBibitem
\bibitem[Zhang \latin{et~al.}(2021)Zhang, Biswas, Shinar, and Shinar]{Zhang:21}
Zhang,~Y.; Biswas,~R.; Shinar,~R.; Shinar,~J. Simulation of enhanced light extraction from periodic, disordered, and quasi-periodic OLED structures. \emph{J. Opt. Soc. Am. B} \textbf{2021}, \emph{38}, C144--C152\relax
\mciteBstWouldAddEndPuncttrue
\mciteSetBstMidEndSepPunct{\mcitedefaultmidpunct}
{\mcitedefaultendpunct}{\mcitedefaultseppunct}\relax
\EndOfBibitem
\bibitem[Kim \latin{et~al.}(2015)Kim, Han, Sung, Kim, Choi, Lee, and Kim]{kim2015enhanced}
Kim,~Y.~D.; Han,~K.-H.; Sung,~Y.~H.; Kim,~J.-B.; Choi,~H.-J.; Lee,~H.; Kim,~J.-J. Enhanced light extraction efficiency in organic light-emitting diode with randomly dispersed nanopattern. \emph{Optics Letters} \textbf{2015}, \emph{40}, 5838--5841\relax
\mciteBstWouldAddEndPuncttrue
\mciteSetBstMidEndSepPunct{\mcitedefaultmidpunct}
{\mcitedefaultendpunct}{\mcitedefaultseppunct}\relax
\EndOfBibitem
\bibitem[Zhu \latin{et~al.}(2016)Zhu, Xiao, Zhai, Yu, Shi, Chen, and Wei]{zhu2016facile}
Zhu,~W.; Xiao,~T.; Zhai,~G.; Yu,~J.; Shi,~G.; Chen,~G.; Wei,~B. A facile method to enhance out-coupling efficiency in organic light-emitting diodes via a random-pyramids textured layer. \emph{Journal of Physics D: Applied Physics} \textbf{2016}, \emph{49}, 385103\relax
\mciteBstWouldAddEndPuncttrue
\mciteSetBstMidEndSepPunct{\mcitedefaultmidpunct}
{\mcitedefaultendpunct}{\mcitedefaultseppunct}\relax
\EndOfBibitem
\bibitem[Kang \latin{et~al.}(2021)Kang, Im, Lee, Kim, and Kim]{kang2021nanoslot}
Kang,~K.; Im,~S.; Lee,~C.; Kim,~J.; Kim,~D. Nanoslot metasurface design and characterization for enhanced organic light-emitting diodes. \emph{Scientific reports} \textbf{2021}, \emph{11}, 9232\relax
\mciteBstWouldAddEndPuncttrue
\mciteSetBstMidEndSepPunct{\mcitedefaultmidpunct}
{\mcitedefaultendpunct}{\mcitedefaultseppunct}\relax
\EndOfBibitem
\bibitem[Kim \latin{et~al.}(2012)Kim, Lee, Kim, Choi, Kweon, Park, and Jeong]{kim2012biologically}
Kim,~J.-J.; Lee,~Y.; Kim,~H.~G.; Choi,~K.-J.; Kweon,~H.-S.; Park,~S.; Jeong,~K.-H. Biologically inspired LED lens from cuticular nanostructures of firefly lantern. \emph{Proceedings of the national academy of sciences} \textbf{2012}, \emph{109}, 18674--18678\relax
\mciteBstWouldAddEndPuncttrue
\mciteSetBstMidEndSepPunct{\mcitedefaultmidpunct}
{\mcitedefaultendpunct}{\mcitedefaultseppunct}\relax
\EndOfBibitem
\bibitem[Greenham \latin{et~al.}(1994)Greenham, Friend, and Bradley]{greenham1994angular}
Greenham,~N.~C.; Friend,~R.~H.; Bradley,~D.~D. Angular dependence of the emission from a conjugated polymer light-emitting diode: implications for efficiency calculations. \emph{Advanced Materials} \textbf{1994}, \emph{6}, 491--494\relax
\mciteBstWouldAddEndPuncttrue
\mciteSetBstMidEndSepPunct{\mcitedefaultmidpunct}
{\mcitedefaultendpunct}{\mcitedefaultseppunct}\relax
\EndOfBibitem
\bibitem[Lee \latin{et~al.}(2003)Lee, Kim, Huh, Kim, Lee, Cho, Kim, and Do]{lee2003high}
Lee,~Y.-J.; Kim,~S.-H.; Huh,~J.; Kim,~G.-H.; Lee,~Y.-H.; Cho,~S.-H.; Kim,~Y.-C.; Do,~Y.~R. A high-extraction-efficiency nanopatterned organic light-emitting diode. \emph{Applied Physics Letters} \textbf{2003}, \emph{82}, 3779--3781\relax
\mciteBstWouldAddEndPuncttrue
\mciteSetBstMidEndSepPunct{\mcitedefaultmidpunct}
{\mcitedefaultendpunct}{\mcitedefaultseppunct}\relax
\EndOfBibitem
\bibitem[Shin \latin{et~al.}(2020)Shin, Huseynova, Kim, Lee, Yoo, Choi, and Lee]{Shin:20}
Shin,~C.-H.; Huseynova,~G.; Kim,~E.; Lee,~J.; Yoo,~S.; Choi,~Y.; Lee,~J.-H. Random Al2O3 nanoparticle-based polymer composite films as outcoupling layers for flexible organic light-emitting diodes. \emph{Opt. Express} \textbf{2020}, \emph{28}, 26170--26179\relax
\mciteBstWouldAddEndPuncttrue
\mciteSetBstMidEndSepPunct{\mcitedefaultmidpunct}
{\mcitedefaultendpunct}{\mcitedefaultseppunct}\relax
\EndOfBibitem
\bibitem[Gu \latin{et~al.}(2013)Gu, Zhang, Ou, Deng, Zhu, Cheng, Liu, Lee, Li, and Tang]{gu2013light}
Gu,~Y.; Zhang,~D.-D.; Ou,~Q.-D.; Deng,~Y.-H.; Zhu,~J.-J.; Cheng,~L.; Liu,~Z.; Lee,~S.-T.; Li,~Y.-Q.; Tang,~J.-X. Light extraction enhancement in organic light-emitting diodes based on localized surface plasmon and light scattering double-effect. \emph{Journal of Materials Chemistry C} \textbf{2013}, \emph{1}, 4319--4326\relax
\mciteBstWouldAddEndPuncttrue
\mciteSetBstMidEndSepPunct{\mcitedefaultmidpunct}
{\mcitedefaultendpunct}{\mcitedefaultseppunct}\relax
\EndOfBibitem
\bibitem[Do \latin{et~al.}(2004)Do, Kim, Song, and Lee]{do2004enhanced}
Do,~Y.~R.; Kim,~Y.-C.; Song,~Y.-W.; Lee,~Y.-H. Enhanced light extraction efficiency from organic light emitting diodes by insertion of a two-dimensional photonic crystal structure. \emph{Journal of Applied Physics} \textbf{2004}, \emph{96}, 7629--7636\relax
\mciteBstWouldAddEndPuncttrue
\mciteSetBstMidEndSepPunct{\mcitedefaultmidpunct}
{\mcitedefaultendpunct}{\mcitedefaultseppunct}\relax
\EndOfBibitem
\bibitem[Cho \latin{et~al.}(2010)Cho, Park, Kim, Jeon, Jeong, and Kim]{cho2010solution}
Cho,~H.-H.; Park,~B.; Kim,~H.-J.; Jeon,~S.; Jeong,~J.-h.; Kim,~J.-J. Solution-processed photonic crystals to enhance the light outcoupling efficiency of organic light-emitting diodes. \emph{Applied optics} \textbf{2010}, \emph{49}, 4024--4028\relax
\mciteBstWouldAddEndPuncttrue
\mciteSetBstMidEndSepPunct{\mcitedefaultmidpunct}
{\mcitedefaultendpunct}{\mcitedefaultseppunct}\relax
\EndOfBibitem
\bibitem[Sung \latin{et~al.}(2017)Sung, Jung, Han, Kim, Kim, and Lee]{sung2017improved}
Sung,~Y.~H.; Jung,~P.-H.; Han,~K.-H.; Kim,~Y.~D.; Kim,~J.-J.; Lee,~H. Improved out-coupling efficiency of organic light emitting diodes fabricated on a TiO2 planarization layer with embedded Si oxide nanostructures. \emph{Optical Materials} \textbf{2017}, \emph{72}, 828--832\relax
\mciteBstWouldAddEndPuncttrue
\mciteSetBstMidEndSepPunct{\mcitedefaultmidpunct}
{\mcitedefaultendpunct}{\mcitedefaultseppunct}\relax
\EndOfBibitem
\bibitem[Ding \latin{et~al.}(2014)Ding, Wang, Chen, and Chou]{ding2014plasmonic}
Ding,~W.; Wang,~Y.; Chen,~H.; Chou,~S.~Y. Plasmonic nanocavity organic light-emitting diode with significantly enhanced light extraction, contrast, viewing angle, brightness, and low-glare. \emph{Advanced Functional Materials} \textbf{2014}, \emph{24}, 6329--6339\relax
\mciteBstWouldAddEndPuncttrue
\mciteSetBstMidEndSepPunct{\mcitedefaultmidpunct}
{\mcitedefaultendpunct}{\mcitedefaultseppunct}\relax
\EndOfBibitem
\bibitem[Wei \latin{et~al.}(2024)Wei, Lin, Xu, Wu, Wang, Wang, Lei, Bao, Li, Li, \latin{et~al.} others]{wei2024inverse}
Wei,~M.; Lin,~X.; Xu,~K.; Wu,~Y.; Wang,~C.; Wang,~Z.; Lei,~K.; Bao,~K.; Li,~J.; Li,~L.; others Inverse design of compact nonvolatile reconfigurable silicon photonic devices with phase-change materials. \emph{Nanophotonics} \textbf{2024}, \emph{13}, 2183--2192\relax
\mciteBstWouldAddEndPuncttrue
\mciteSetBstMidEndSepPunct{\mcitedefaultmidpunct}
{\mcitedefaultendpunct}{\mcitedefaultseppunct}\relax
\EndOfBibitem
\bibitem[Minkov \latin{et~al.}(2020)Minkov, Williamson, Andreani, Gerace, Lou, Song, Hughes, and Fan]{minkov2020inverse}
Minkov,~M.; Williamson,~I.~A.; Andreani,~L.~C.; Gerace,~D.; Lou,~B.; Song,~A.~Y.; Hughes,~T.~W.; Fan,~S. Inverse design of photonic crystals through automatic differentiation. \emph{Acs Photonics} \textbf{2020}, \emph{7}, 1729--1741\relax
\mciteBstWouldAddEndPuncttrue
\mciteSetBstMidEndSepPunct{\mcitedefaultmidpunct}
{\mcitedefaultendpunct}{\mcitedefaultseppunct}\relax
\EndOfBibitem
\bibitem[Maier \latin{et~al.}(2007)Maier, \latin{et~al.} others]{maier2007plasmonics}
Maier,~S.~A.; others \emph{Plasmonics: fundamentals and applications}; Springer, 2007; Vol.~1\relax
\mciteBstWouldAddEndPuncttrue
\mciteSetBstMidEndSepPunct{\mcitedefaultmidpunct}
{\mcitedefaultendpunct}{\mcitedefaultseppunct}\relax
\EndOfBibitem
\bibitem[Benson(2012)]{benson2012fields}
Benson,~M. \emph{Fields, waves and transmission lines}; Springer Science \& Business Media, 2012\relax
\mciteBstWouldAddEndPuncttrue
\mciteSetBstMidEndSepPunct{\mcitedefaultmidpunct}
{\mcitedefaultendpunct}{\mcitedefaultseppunct}\relax
\EndOfBibitem
\bibitem[Liao(1990)]{liao1990microwave}
Liao,~S.~Y. \emph{Microwave devices and circuits}; Pearson Education India, 1990\relax
\mciteBstWouldAddEndPuncttrue
\mciteSetBstMidEndSepPunct{\mcitedefaultmidpunct}
{\mcitedefaultendpunct}{\mcitedefaultseppunct}\relax
\EndOfBibitem
\bibitem[Jeong \latin{et~al.}(2015)Jeong, Salas-Montiel, and Jeong]{Jeong:15}
Jeong,~H.; Salas-Montiel,~R.; Jeong,~M.~S. Optimal length of ZnO nanorods for improving the light-extraction efficiency of blue InGaN light-emitting diodes. \emph{Opt. Express} \textbf{2015}, \emph{23}, 23195--23207\relax
\mciteBstWouldAddEndPuncttrue
\mciteSetBstMidEndSepPunct{\mcitedefaultmidpunct}
{\mcitedefaultendpunct}{\mcitedefaultseppunct}\relax
\EndOfBibitem
\bibitem[Lu \latin{et~al.}(2016)Lu, Shi, Wang, Lin, Zhu, Tian, Dai, Wang, and Xu]{lu2016plasmon}
Lu,~J.; Shi,~Z.; Wang,~Y.; Lin,~Y.; Zhu,~Q.; Tian,~Z.; Dai,~J.; Wang,~S.; Xu,~C. Plasmon-enhanced electrically light-emitting from ZnO nanorod arrays/p-GaN heterostructure devices. \emph{Scientific Reports} \textbf{2016}, \emph{6}, 25645\relax
\mciteBstWouldAddEndPuncttrue
\mciteSetBstMidEndSepPunct{\mcitedefaultmidpunct}
{\mcitedefaultendpunct}{\mcitedefaultseppunct}\relax
\EndOfBibitem
\bibitem[Mangalgiri \latin{et~al.}(2017)Mangalgiri, Manley, Riedel, and Schmid]{mangalgiri2017dielectric}
Mangalgiri,~G.~M.; Manley,~P.; Riedel,~W.; Schmid,~M. Dielectric nanorod scattering and its influence on material interfaces. \emph{Scientific Reports} \textbf{2017}, \emph{7}, 4311\relax
\mciteBstWouldAddEndPuncttrue
\mciteSetBstMidEndSepPunct{\mcitedefaultmidpunct}
{\mcitedefaultendpunct}{\mcitedefaultseppunct}\relax
\EndOfBibitem
\bibitem[Van~de Groep and Polman(2013)Van~de Groep, and Polman]{van2013designing}
Van~de Groep,~J.; Polman,~A. Designing dielectric resonators on substrates: Combining magnetic and electric resonances. \emph{Optics express} \textbf{2013}, \emph{21}, 26285--26302\relax
\mciteBstWouldAddEndPuncttrue
\mciteSetBstMidEndSepPunct{\mcitedefaultmidpunct}
{\mcitedefaultendpunct}{\mcitedefaultseppunct}\relax
\EndOfBibitem
\bibitem[Hsiao \latin{et~al.}(2014)Hsiao, Chen, Huang, Lin, Lien, Huang, and He]{Hsiao2014}
Hsiao,~Y.-H.; Chen,~C.-Y.; Huang,~L.-C.; Lin,~G.-J.; Lien,~D.-H.; Huang,~J.-J.; He,~J.-H. Light Extraction Enhancement with Radiation Pattern Shaping of LEDs by Waveguiding Nanorods with Impedance-Matching Tips. \emph{Nanoscale} \textbf{2014}, \emph{6}, 2624--2628\relax
\mciteBstWouldAddEndPuncttrue
\mciteSetBstMidEndSepPunct{\mcitedefaultmidpunct}
{\mcitedefaultendpunct}{\mcitedefaultseppunct}\relax
\EndOfBibitem
\bibitem[Kim and Osterloh(2005)Kim, and Osterloh]{Kim2005}
Kim,~J.~Y.; Osterloh,~F.~E. ZnO--CdSe Nanoparticle Clusters as Directional Photoemitters with Tunable Wavelength. \emph{Journal of the American Chemical Society} \textbf{2005}, \emph{127}, 10152--10153\relax
\mciteBstWouldAddEndPuncttrue
\mciteSetBstMidEndSepPunct{\mcitedefaultmidpunct}
{\mcitedefaultendpunct}{\mcitedefaultseppunct}\relax
\EndOfBibitem
\bibitem[Kwon \latin{et~al.}(2018)Kwon, Jang, Park, Kim, Hong, Jung, Yang, Kim, and Cho]{kwon2018plasmonic}
Kwon,~O.~H.; Jang,~J.~W.; Park,~S.-J.; Kim,~J.~S.; Hong,~S.~J.; Jung,~Y.~S.; Yang,~H.; Kim,~Y.~J.; Cho,~Y.~S. Plasmonic-enhanced luminescence characteristics of microscale phosphor layers on a ZnO nanorod-arrayed glass substrate. \emph{ACS Applied Materials \& Interfaces} \textbf{2018}, \emph{11}, 1004--1012\relax
\mciteBstWouldAddEndPuncttrue
\mciteSetBstMidEndSepPunct{\mcitedefaultmidpunct}
{\mcitedefaultendpunct}{\mcitedefaultseppunct}\relax
\EndOfBibitem
\bibitem[Xu \latin{et~al.}(2019)Xu, Hong, and Kuo]{https://doi.org/10.1002/sdtp.13343}
Xu,~W.-F.; Hong,~C.-T.; Kuo,~C.-K. P-186: Reducing Angular Color Shift in RGB OLEDs by the Introduction of a Circular Polarizer with a Diffractive Optical Element. \emph{SID Symposium Digest of Technical Papers} \textbf{2019}, \emph{50}, 1932--1934\relax
\mciteBstWouldAddEndPuncttrue
\mciteSetBstMidEndSepPunct{\mcitedefaultmidpunct}
{\mcitedefaultendpunct}{\mcitedefaultseppunct}\relax
\EndOfBibitem
\bibitem[Siddiqui \latin{et~al.}(2023)Siddiqui, Kumar, Tsai, Gautam, Shahnawaz, Kesavan, Lin, Khai, Chou, Choudhury, Grigalevicius, and Jou]{nano13182521}
Siddiqui,~I.; Kumar,~S.; Tsai,~Y.-F.; Gautam,~P.; Shahnawaz; Kesavan,~K.; Lin,~J.-T.; Khai,~L.; Chou,~K.-H.; Choudhury,~A.; Grigalevicius,~S.; Jou,~J.-H. Status and Challenges of Blue OLEDs: A Review. \emph{Nanomaterials} \textbf{2023}, \emph{13}\relax
\mciteBstWouldAddEndPuncttrue
\mciteSetBstMidEndSepPunct{\mcitedefaultmidpunct}
{\mcitedefaultendpunct}{\mcitedefaultseppunct}\relax
\EndOfBibitem
\bibitem[Chen \latin{et~al.}(2025)Chen, Gu, Guo, Xu, Shi, Li, Cao, and Liu]{Chen:25}
Chen,~L.; Gu,~H.; Guo,~X.; Xu,~M.; Shi,~T.; Li,~J.; Cao,~W.; Liu,~S. Simultaneously improving color purity, stability, and health-friendliness of OLEDs via optimally designed color filters. \emph{Opt. Express} \textbf{2025}, \emph{33}, 13506--13518\relax
\mciteBstWouldAddEndPuncttrue
\mciteSetBstMidEndSepPunct{\mcitedefaultmidpunct}
{\mcitedefaultendpunct}{\mcitedefaultseppunct}\relax
\EndOfBibitem
\bibitem[Sun and Tsui(2007)Sun, and Tsui]{sun2007production}
Sun,~Y.; Tsui,~Y. Production of porous nanostructured zinc oxide thin films by pulsed laser deposition. \emph{Optical materials} \textbf{2007}, \emph{29}, 1111--1114\relax
\mciteBstWouldAddEndPuncttrue
\mciteSetBstMidEndSepPunct{\mcitedefaultmidpunct}
{\mcitedefaultendpunct}{\mcitedefaultseppunct}\relax
\EndOfBibitem
\bibitem[Taschuk \latin{et~al.}(2008)Taschuk, Sun, and Tsui]{taschuk2008angularly}
Taschuk,~M.; Sun,~Y.; Tsui,~Y. Angularly resolved photoluminescent emission from pulsed-laser-deposited ZnO films with different microstructures. \emph{Applied Physics A} \textbf{2008}, \emph{90}, 141--147\relax
\mciteBstWouldAddEndPuncttrue
\mciteSetBstMidEndSepPunct{\mcitedefaultmidpunct}
{\mcitedefaultendpunct}{\mcitedefaultseppunct}\relax
\EndOfBibitem
\bibitem[Thandavan \latin{et~al.}(2015)Thandavan, Gani, San~Wong, and Md.~Nor]{thandavan2015enhanced}
Thandavan,~T. M.~K.; Gani,~S. M.~A.; San~Wong,~C.; Md.~Nor,~R. Enhanced photoluminescence and Raman properties of Al-doped ZnO nanostructures prepared using thermal chemical vapor deposition of methanol assisted with heated brass. \emph{PLoS One} \textbf{2015}, \emph{10}, e0121756\relax
\mciteBstWouldAddEndPuncttrue
\mciteSetBstMidEndSepPunct{\mcitedefaultmidpunct}
{\mcitedefaultendpunct}{\mcitedefaultseppunct}\relax
\EndOfBibitem
\bibitem[Sun \latin{et~al.}(2013)Sun, Jiao, Zhang, Wang, Li, Gao, Wang, Yu, Guo, Zhao, \latin{et~al.} others]{sun2013morphology}
Sun,~S.; Jiao,~S.; Zhang,~K.; Wang,~D.; Li,~H.; Gao,~S.; Wang,~J.; Yu,~Q.; Guo,~F.; Zhao,~L.; others Morphology and properties of ZnO nanostructures by electrochemical deposition: effect of the substrate treatment. \emph{Journal of Materials Science: Materials in Electronics} \textbf{2013}, \emph{24}, 85--88\relax
\mciteBstWouldAddEndPuncttrue
\mciteSetBstMidEndSepPunct{\mcitedefaultmidpunct}
{\mcitedefaultendpunct}{\mcitedefaultseppunct}\relax
\EndOfBibitem
\bibitem[Dong \latin{et~al.}(2012)Dong, Han, Qian, and Chen]{dong2012fluorescent}
Dong,~Z.; Han,~B.; Qian,~S.; Chen,~D. Fluorescent properties of ZnO nanostructures fabricated by hydrothermal method. \emph{Journal of Nanomaterials} \textbf{2012}, \emph{2012}, 251276\relax
\mciteBstWouldAddEndPuncttrue
\mciteSetBstMidEndSepPunct{\mcitedefaultmidpunct}
{\mcitedefaultendpunct}{\mcitedefaultseppunct}\relax
\EndOfBibitem
\bibitem[Nandanapalli and Mudusu(2018)Nandanapalli, and Mudusu]{nandanapalli2018surface}
Nandanapalli,~K.~R.; Mudusu,~D. Surface passivated zinc oxide (ZnO) nanorods by atomic layer deposition of ultrathin ZnO layers for energy device applications. \emph{ACS Applied Nano Materials} \textbf{2018}, \emph{1}, 4083--4091\relax
\mciteBstWouldAddEndPuncttrue
\mciteSetBstMidEndSepPunct{\mcitedefaultmidpunct}
{\mcitedefaultendpunct}{\mcitedefaultseppunct}\relax
\EndOfBibitem
\bibitem[Zheng \latin{et~al.}(2024)Zheng, Rosli, Rashid, and Halim]{zheng2024electrically}
Zheng,~K.~O.; Rosli,~N.; Rashid,~M.; Halim,~M.~M. Electrically pumped random laser device based on Pd/SiO2/ZnO nanorods MIS structure. \emph{Results in Physics} \textbf{2024}, \emph{64}, 107946\relax
\mciteBstWouldAddEndPuncttrue
\mciteSetBstMidEndSepPunct{\mcitedefaultmidpunct}
{\mcitedefaultendpunct}{\mcitedefaultseppunct}\relax
\EndOfBibitem
\bibitem[Hackett \latin{et~al.}(2018)Hackett, Ameen, Li, Dar, Goddard, and Liu]{hackett2018spectrometer}
Hackett,~L.~P.; Ameen,~A.; Li,~W.; Dar,~F.~K.; Goddard,~L.~L.; Liu,~G.~L. Spectrometer-free plasmonic biosensing with metal--insulator--metal nanocup arrays. \emph{ACS sensors} \textbf{2018}, \emph{3}, 290--298\relax
\mciteBstWouldAddEndPuncttrue
\mciteSetBstMidEndSepPunct{\mcitedefaultmidpunct}
{\mcitedefaultendpunct}{\mcitedefaultseppunct}\relax
\EndOfBibitem
\end{mcitethebibliography}
\IfFileExists{achemso-demo.bbl}{\providecommand{\latin}[1]{#1}
\makeatletter
\providecommand{\doi}
  {\begingroup\let\do\@makeother\dospecials
  \catcode`\{=1 \catcode`\}=2 \doi@aux}
\providecommand{\doi@aux}[1]{\endgroup\texttt{#1}}
\makeatother
\providecommand*\mcitethebibliography{\thebibliography}
\csname @ifundefined\endcsname{endmcitethebibliography}  {\let\endmcitethebibliography\endthebibliography}{}

}{}

\end{document}